\documentclass[showpacs,aps,prb,reprint,superscriptaddress,preprintnumbers]{revtex4-2}

\usepackage{amsmath}
\usepackage{amssymb}
\usepackage{graphicx}% Include figure files
\usepackage{dcolumn}
\usepackage{color}
\usepackage{xcolor}
\usepackage{endnotes}
\usepackage{bm}
\usepackage{epsfig}
\usepackage{array}
\usepackage{float}
\usepackage{tabularx}

\newcommand{\etal}{{\it et al.}}
\usepackage{soul}

\begin{document}

%\preprint{v3.1}

\title{Suppression of both superconductivity and structural transition in hole-doped MoTe$_2$ induced by Ta substitution}

\author{Siu Tung Lam}
\author{K.~Y.~Yip}
\author{Swee K. Goh}
\email[]{skgoh@cuhk.edu.hk}
\affiliation{Department of Physics, The Chinese University of Hong Kong, Shatin, Hong Kong, China}
\author{Kwing To Lai}
\email[]{ktlai@phy.cuhk.edu.hk}
\affiliation{Department of Physics, The Chinese University of Hong Kong, Shatin, Hong Kong, China} 
\affiliation{Shenzhen Research Institute, The Chinese University of Hong Kong, Shatin, Hong Kong, China}
\affiliation{Faculty of Science, The University of Hong Kong, Pokfulam Road, Hong Kong, China}

\date{\today}

\begin{abstract}

  Type-II Weyl semimetal MoTe$_2$ exhibits a first-order structural transition at $T_s$ $\sim$250~K and superconducts at $T_c$ $\sim$0.1~K at ambient pressure. Both $T_s$ and $T_c$ can be manipulated by several tuning parameters, such as hydrostatic pressure and chemical substitution. It is often reported that suppressing $T_s$ enhances $T_c$, but our study shows a different behaviour when MoTe$_2$ is hole-doped by Ta. When $T_s$ is suppressed by Ta doping, $T_c$ is also suppressed. Our findings suggest that the suppression of $T_s$ does not necessarily enhance superconductivity in MoTe$_2$. By connecting with the findings of electron-doped MoTe$_2$, we argue that varying electron carrier concentration can effectively tune $T_c$. In addition, the Hall coefficient is enhanced around the doping region, where $T_s$ is completely suppressed, suggesting that the critical scattering around the structural transition may also play a role in suppressing $T_c$.
\end{abstract}

%\pacs{Preliminary draft, not all references included}
%\keywords{}

\maketitle
\section{Introduction}
Superconductivity is found, often by tuning the electronic properties via the application of hydrostatic pressure, in many topological semimetals, such as Cd$_3$As$_2$ \cite{He2016qm}, ZrTe$_5$ \cite{Zhou2016pnas}, YPtBi \cite{Meinert2016prl,Butch2011prbrc,Bay2012,Kim2018scnadv}, WTe$_2$ \cite{Pan2015,Kang2015,Lu2016,Chan2017} and MoTe$_2$ \cite{Qi2016,Rhodes2017,Hu2019,Hu2020}. The exotic combination of topological bands and superconductivity offers a unique platform to search for topological superconductivity, where Majorana fermions can be used to develop topological quantum computation \cite{Sato2017,Li2019aqt}. 

Type-II Weyl semimetal MoTe$_2$ \cite{Soluyanov2015,Deng2016,Huang2016,Jiang2017nc} is one of the promising candidates for hosting topological superconductivity, especially after the discovery of an edge supercurrent \cite{Wang2020}. At ambient pressure, MoTe$_2$ undergoes a first-order structural transition at $T_s \sim$ 250~K, changing from a centrosymmetric (nonpolar) monoclinic $1T'$ phase (space group: $P2_1/m$) to a noncentrosymmetric (polar) orthorhombic $T_d$ phase (space group: $Pmn2_1$) upon cooling. At $T_c \sim$ 0.1~K, an additional superconducting phase transition occurs.

Owing to its low $T_c$, it is challenging to experimentally study the superconductivity of MoTe$_2$. Finding a suitable way to control its $T_c$ becomes an outstanding issue. Meanwhile, the competition between structural and superconducting transitions in MoTe$_2$ has been reported in previous studies using a variety of tuning parameters. Through the application of pressure \cite{Qi2016,Takahashi2017,Heikes2018,Lee2018,Guguchia2017,Hu2019}, $T_s$ is suppressed to 0~K at $\sim$10~kbar, resulting in a complete removal of the $T_d$ phase at high pressures. Meanwhile, $T_c$ is enhanced by 30-fold ($\sim$4~K) at $\sim$15~kbar. These behaviours demonstrate the anticorrelation between $T_s$ and $T_c$. Similar anticorrelation can also be observed via isovalent chemical substitutions (S/Se substituting Te \cite{Takahashi2017,Chen2016}) and electron doping (Te deficiency \cite{Cho2017} and Re substituting Mo \cite{Mandal2018}). Note that via the substitution of Mo by W, $T_s$ is enhanced at ambient pressure, and the pressure-induced $T_c$ is lower than that observed in the pristine MoTe$_2$, demonstrating the anticorrelation between $T_s$ and $T_c$ again~\cite{Dahal2020}. 

The superconductivity of hole-doped MoTe$_2$ has not been studied to the same extent as the electron-doped counterpart. The introduction of hole carriers in monolayer MoTe$_2$ through gating has been shown to reduce its $T_c$ \cite{Rhodes2021}. On the other hand, the effect of hole doping on bulk MoTe$_2$ has been explored in depth through the substitution of Nb for Mo \cite{Ikeura2015,Sakai2016}. Although no evidence of superconductivity with $T_c >$~2~K has been found up to the highest studied doping level $x$ = 0.22 in Mo$_{1-x}$Nb$_x$Te$_2$, indicating a lack of significant enhancement in $T_c$ through hole doping, the hole-doping phase diagram of Mo$_{1-x}$Nb$_x$Te$_2$ in the normal state was extensively investigated by Sakai \etal~\cite{Sakai2016}. They revealed that the suppression of $T_s$ upon Nb doping is associated with a huge enhancement of thermopower at low temperatures, which they attributed to the critical scattering arising from the boundary of the nonpolar-to-polar transition around $T_s$.

Nevertheless, it remains uncertain how $T_c$ evolves and what the correlation of $T_s$ and $T_c$ is upon hole doping. Understanding these issues can help us reveal the key factors that control $T_c$ of MoTe$_2$. In this article, we study the effect of hole doping on MoTe$_2$ via the substitution of Mo by Ta. Transport measurements were conducted down to $\sim$30 mK to track the evolution of both $T_s$ and $T_c$, and surprisingly, we found that both $T_s$ and $T_c$ are suppressed and eventually vanish with increasing hole doping, contrary to the anticorrelation between $T_s$ and $T_c$ established in MoTe$_2$ controlled by other tuning parameters.

%%%%%%%%%%%%%%%%Figure 1
\begin{figure}[!t]\centering
      \resizebox{8.8cm}{!}{
              \includegraphics{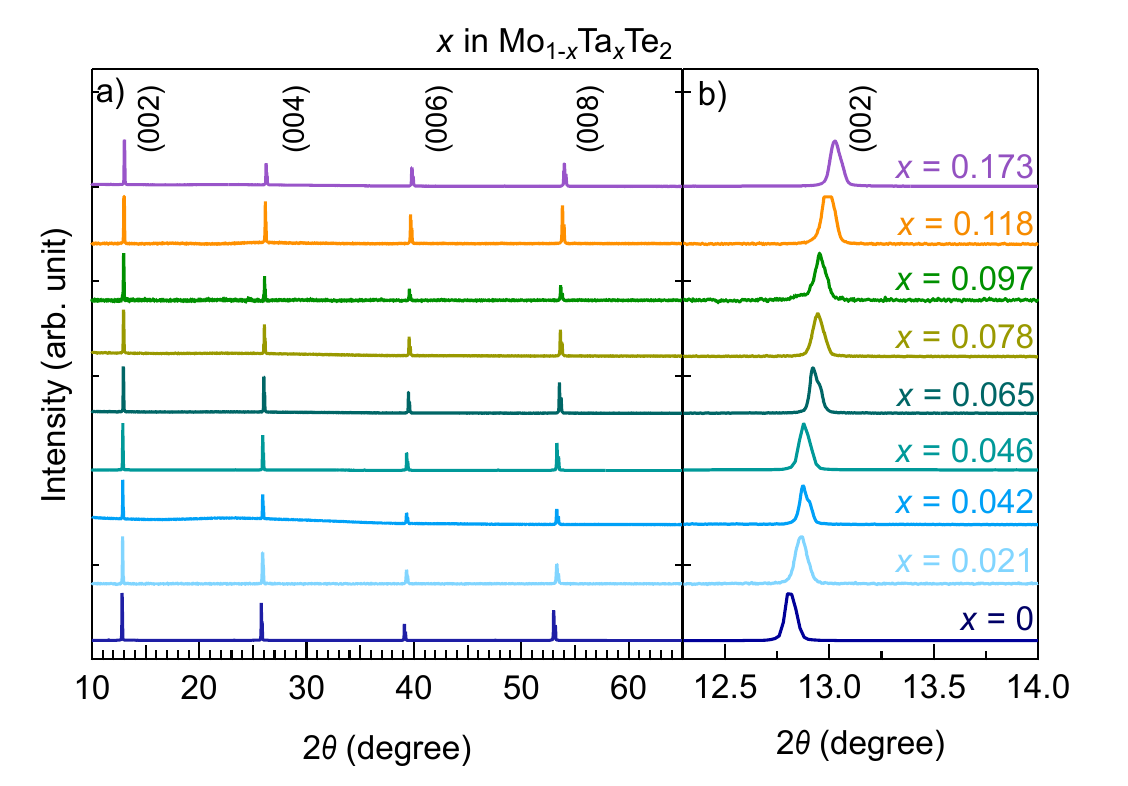}}                				
              \caption{\label{fig1} (a) X-ray diffraction (XRD) spectra of single crystals of Mo$_{1-x}$Ta$_x$Te$_2$. The peaks of (00$L$) are indexed in the figure. (b) Enlarged XRD spectra near the peak of (002). The (002) peak shifts progressively toward a higher diffraction angle $2\theta$ when $x$ increases.
              }
\end{figure}
%%%%%%%%%%%%%%%%%%%%%

\section{Experiment}

Single crystals of Mo$_{1-x}$Ta$_x$Te$_2$ were grown by the self-flux method. The mixture of Mo powder (99.999$\%$, Alfa Aesar), Te (99.99999$\%$ lumps, Ultimate Material), and Ta powder (99.99$\%$, Sigma Aldrich) were first placed into an alumina crucible, with a stoichiometric ratio of Mo:Ta:Te = 1$-x$:$x$:20. The alumina crucible was inserted into a quartz tube before the quartz tube was sealed under a vacuum. The sealed ampule was then heated to 1100~$^\circ$C within 24 hours and stayed for 24 hours, followed by slow cooling to 880~$^\circ$C for 400 hours. Finally, the ampule was taken out from the furnace at 880~$^\circ$C and centrifuged to remove the excess Te flux. X-ray diffraction (XRD) data were collected at room temperature by using a Rigaku X-ray diffractometer with Cu$K_\alpha$ radiation. The chemical compositions were characterized by a JEOL JSM-7800F scanning electron microscope equipped with an Oxford energy-dispersive X-ray (EDX) spectrometer. A standard four-probe method was used to measure temperature-dependent resistance in a Bluefors dilution refrigerator with a base temperature of 30~mK. A standard six-probe method was used to measure the Hall effect in a Quantum Design Physical Property Measurement System with a temperature range from 300~K to 2~K and a magnetic field of $\pm14$~T.

\section{Results and discussion}
Figure~\ref{fig1}(a) shows the XRD spectra for the Mo$_{1-x}$Ta$_x$Te$_2$ single crystals with $x=$ 0, 0.021, 0.042, 0.046, 0.065, 0.078, 0.097, 0.118, and 0.173. The peaks shown in all spectra are well indexed by the (00$L$) planes originating from the pattern of $1T'$-MoTe$_2$, confirming that all crystals are single-crystalline $1T'$-MoTe$_2$ at room temperature. Figure~\ref{fig1}(b) focuses on the (002) peaks of all samples, which reveal a monotonic shift to a higher 2$\theta$ when $x$ increases, indicating a shrinking crystal structure. As the covalent radius of Ta is smaller than that of Mo, this provides crystallographic evidence that Ta is systemically substituting Mo with increasing $x$. These Mo$_{1-x}$Ta$_x$Te$_2$ crystals measured in XRD were also examined by EDX, from which we determined their elemental compositions and hence the values of $x$ in each sample. The EDX results are consistent with the findings in XRD spectra. (see Supplemental Material for more details \cite{supp}.)

%%%%%%%%%%%%%%%%Figure 2
\begin{figure}[!t]\centering
       \resizebox{7cm}{!}{
              \includegraphics{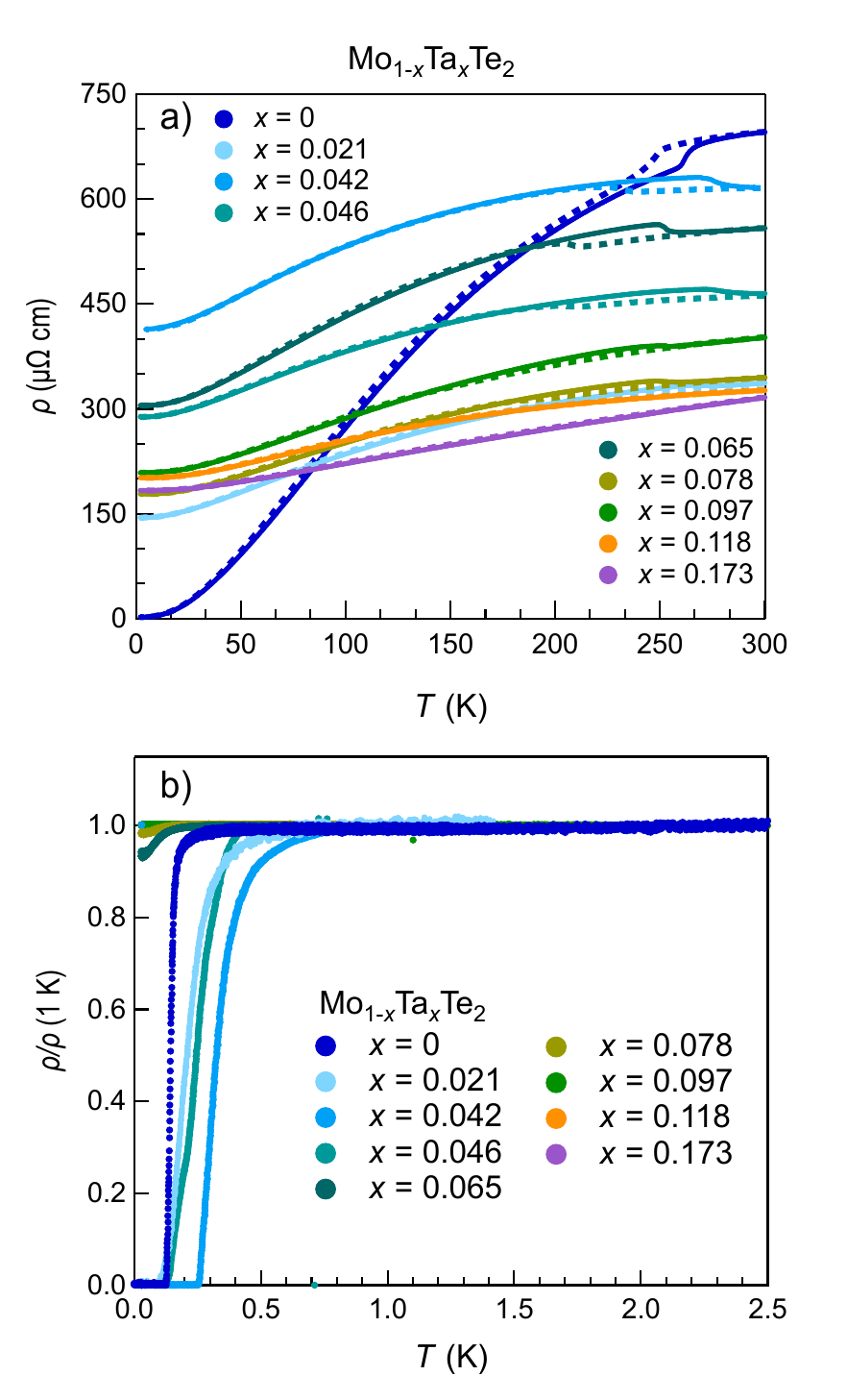}}  
              \caption{\label{fig2} (a) Temperature dependence of resistivity $\rho(T)$ of Mo$_{1-x}$Ta$_x$Te$_2$ at zero magnetic fields. The warm-up (cool-down) data are plotted as solid (dashed) curves. (b) Low-temperature $\rho(T)$ normalized to the value of $\rho(1~\mathrm{K})$, displaying the superconducting transitions.
               }
\end{figure}
%%%%%%%%%%%%%%%%%%%%

Figure~\ref{fig2}(a) illustrates the temperature dependence of resistivity $\rho(T)$ of Mo$_{1-x}$Ta$_x$Te$_2$ with $x=0-0.173$ measured under zero magnetic field. All samples exhibit metallic behaviour. A thermal hysteresis can be observed in pristine MoTe$_2$ ($x=0$) around 150--250~K when the resistivity was measured upon increasing (solid curves) and decreasing temperature (dashed curves), indicating the appearance of the first-order structural transition \cite{Zandt2007,Qi2016,Hu2019,Hu2020,Yip2023}. This transition persists up to $x=0.097$. With increasing $x$, the transition shifts gradually toward lower temperatures, and the hysteresis loop becomes broader. When $x\geq 0.118$, no hysteresis is observed in the whole temperature range, suggesting that the structural transition vanishes at the high doping region.
Figure~\ref{fig2}(b) shows the resistivity data normalized to the value of $\rho(T)$ at 1~K at the low-temperature region. A superconducting transition, where $T_c$ is defined at which the resistivity drops to zero, is observed at $x=0$ with $T_c \sim$ 0.1~K, which is consistent with the previous studies \cite{Qi2016,Rhodes2017,Wang2020,Hu2019,Hu2020,Takahashi2017,Heikes2018,Lee2018,Guguchia2017,Yip2023}. When $x$ increases, $T_c$ generally reduces despite a small enhancement to $\sim$0.25~K at $x=$~0.042. At $x=$~0.065, a small drop of resistivity without reaching zero resistivity is observed near the base temperature, indicating that the bulk superconductivity is heavily suppressed and only trace superconductivity is detected. When $x$ further increases ($\geq 0.078$), the resistivity data shows no signs of superconductivity.

\begin{figure}[!t]\centering
       \resizebox{8.2cm}{!}{
              \includegraphics{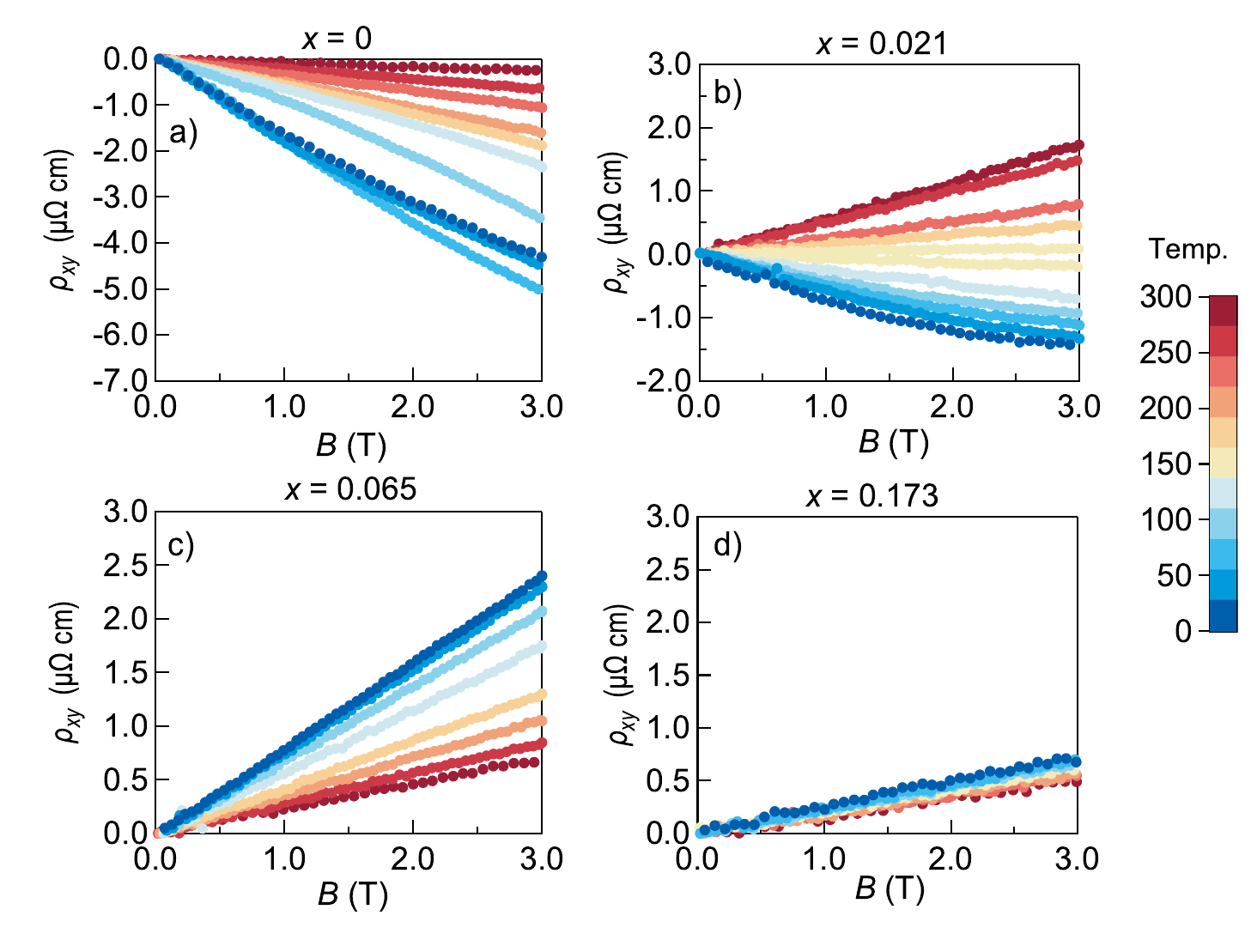}}                				
              \caption{\label{fig3} Magnetic field dependence of Hall resistivity  $\rho_{xy}(B)$ of Mo$_{1-x}$Ta$_x$Te$_2$ with (a) $x=0$, (b) $x=0.021$, (c) $x=0.065$, and (d) $x=0.173$ collected during warm-up. The colour scale at the right indicates the measured temperature. 
              }
\end{figure}

To probe the evolution of the Fermi surface of Mo$_{1-x}$Ta$_x$Te$_2$, we conducted the Hall effect measurements. Figure~\ref{fig3} illustrates the magnetic field dependence of Hall resistivity $\rho_{xy}(B)$ of Mo$_{1-x}$Ta$_x$Te$_2$ with $x=0$, 0.021, 0.065 and 0.173 at different temperatures measured during warm-up. $\rho_{xy}(B)$ data of samples with other doping can be found in Fig.~S2 in Supplemental Material \cite{supp}. At $x=0$ (Fig.~\ref{fig3}(a)), $\rho_{xy}(B)$ has a negative slope at the whole temperature range. At low temperatures, $\rho_{xy}(B)$ shows a non-linear feature. These features are consistent with the semimetallic nature of MoTe$_2$, which exhibits nearly perfect electron-hole compensation with a high electron mobility \cite{Zhou2016Hall,Hu2019}. After introducing Ta doping, the slope of $\rho_{xy}(B)$ at $x=0.021$ (Fig.~\ref{fig3}(b)) begins to turn positive at high temperatures. When $x$ further increases, the slope is always positive at all measured temperatures (see Figs.~\ref{fig3}(c) and (d) as examples). This trend indicates that Ta doping introduces hole carriers to the samples and the hole carriers are dominant at $x>0.021$. Moreover, the additional hole carriers destroy the nearly perfect electron-hole compensation, resulting in the linear positive slope of $\rho_{xy}(B)$ at $x>0.021$.

%%%%%%%%%%%%%%%%Figure 3
%%%%%%%%%%%%%%%%%%%%

\begin{figure}[!t]\centering
       \resizebox{8cm}{!}{
              \includegraphics{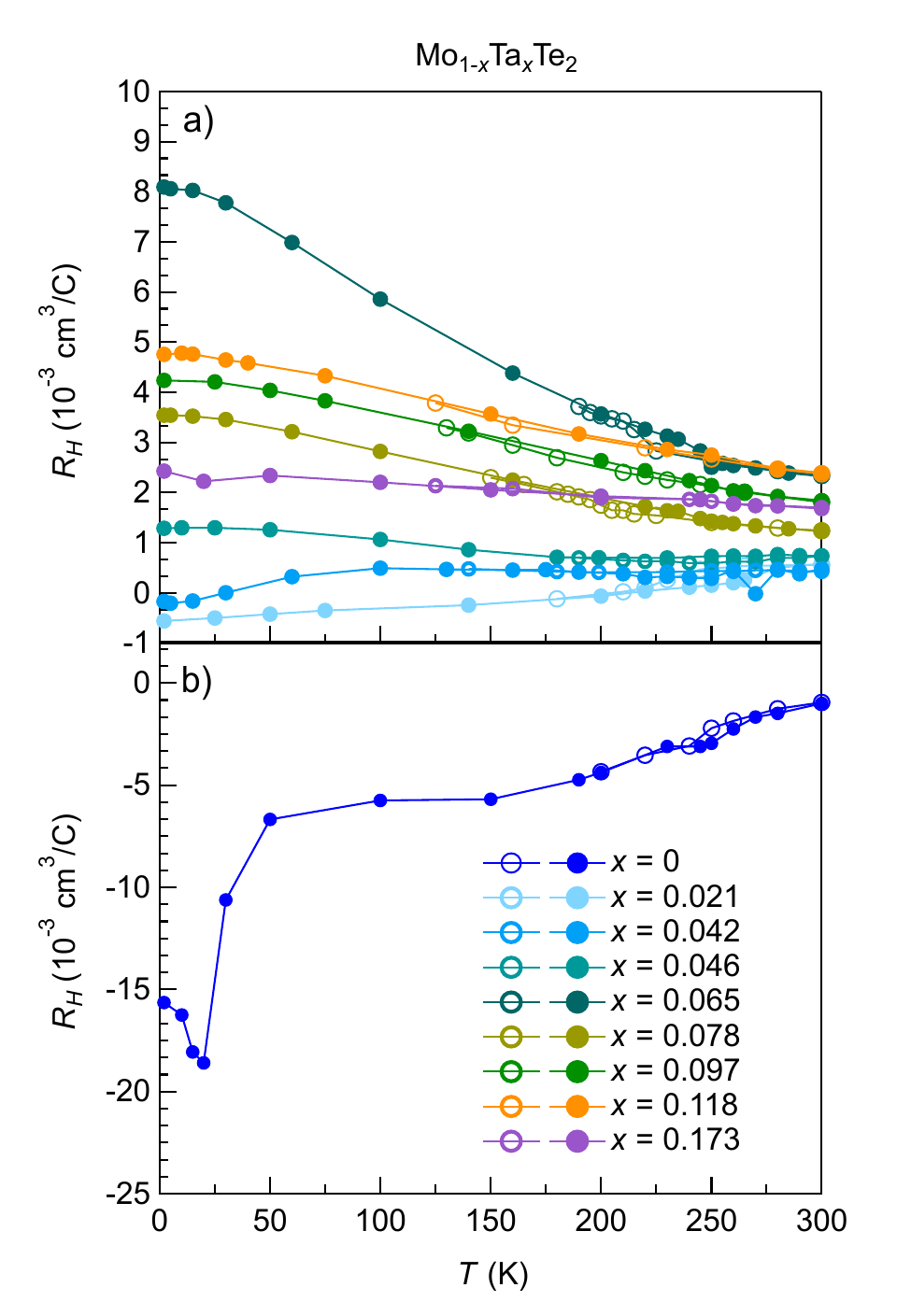}}                				
              \caption{\label{fig4} Temperature dependence of Hall coefficient $R_H$ of (a) Mo$_{1-x}$Ta$_x$Te$_2$ with $x\neq 0$ and (b) pristine MoTe$_2$ ($x=0$). The closed (open) symbols represent the warm-up (cool-down) data. The cool-down data are only shown at high temperatures.
              }
\end{figure}

To further visualize the temperature evolution of the Hall effect of Mo$_{1-x}$Ta$_x$Te$_2$, we extract the Hall coefficient $R_H$ from the slope of $\rho_{xy}(B)$ in the linear region, and the temperature evolution of $R_H$ is plotted in Fig.~\ref{fig4}. The $R_H$ data measured at high temperatures during cool-down are also displayed. We find that a thermal hysteresis can also be observed in the $R_H$ data of the samples from $x=0$ to $x=0.097$, while the hysteresis is absent in the sample with $x\geq 0.118$. These results are consistent with the observation of the first-order structural transition in the $\rho(T)$ data in Fig.~\ref{fig2}(a). At $x=0$ (Fig.~\ref{fig4}(b)), $R_H$ shows a strong temperature dependence below 50~K, which is similar to the result reported in previous studies \cite{Zhou2016Hall,Zandt2007}. Upon Ta doping, $R_H$ shifts toward the positive side due to the introduction of additional hole carriers, while the temperature dependence is relatively mild compared to $x=0$. The most prominent temperature profile of $R_H$ in Ta-doped samples is $x=0.065$, where the magnitude of $R_H$ ($\vert R_H \vert$) gradually increases with decreasing temperature and reaches the maximum value at 2~K. Interestingly, our results show that $\vert R_H \vert$ at 2~K is the largest around $x=0.065$ (see also Fig.~\ref{fig3}(c), where $\rho_{xy}(B)$ has a steeper slope at 2~K compared to that in Fig.~\ref{fig3}(d)), which is different from the expectation that $R_H$ would increase toward the positive side when $x$ increases. This issue will be further discussed in the later section.

%%%%%%%%%%%%%%%%Figure 4
%%%%%%%%%%%%%%%%%%%%

\begin{figure}[!t]\centering
       \resizebox{8cm}{!}{
              \includegraphics{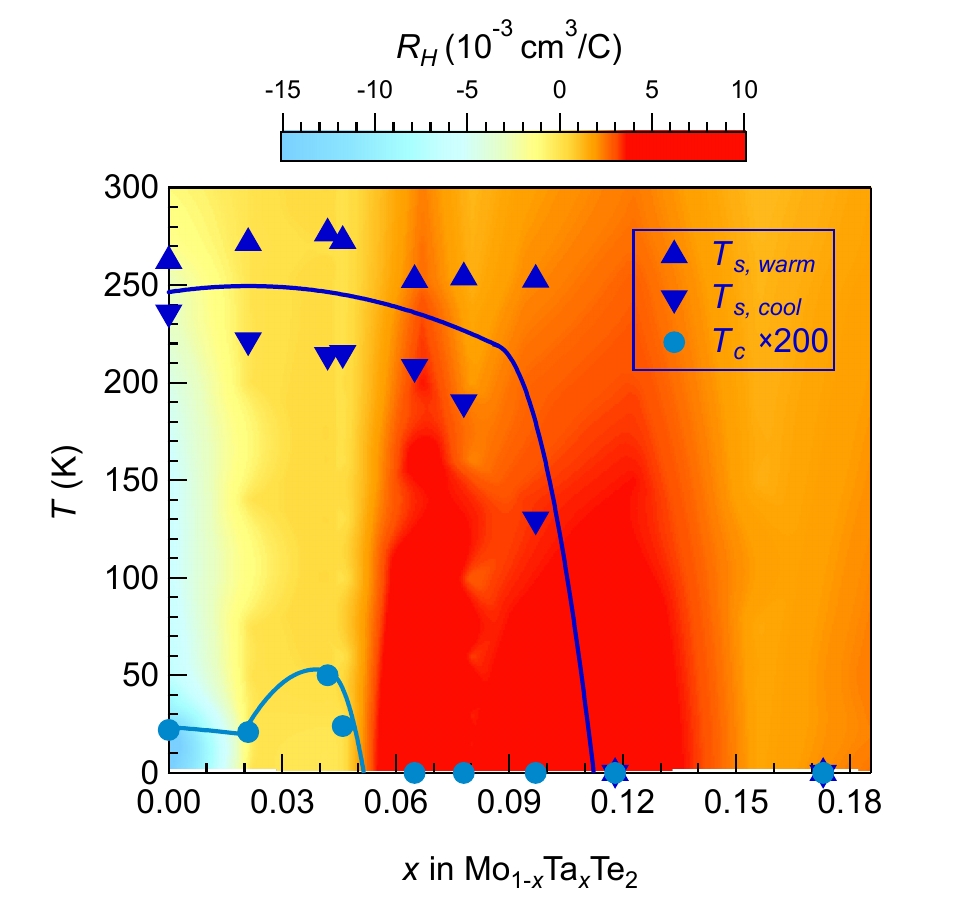}}                				
              \caption{\label{fig5} Temperature-doping phase diagram of Mo$_{1-x}$Ta$_x$Te$_2$. The upward (downward) blue triangles represent $T_s$ defined from the temperature-dependent resistivity data measured during warm-up (cool-down). The solid cyan circles represent $T_c$. The solid curves are guides for the eyes. The colour contour denotes the temperature dependence of Hall coefficient $R_H$ at different doping levels.
              }
\end{figure}

We summarize our results and construct a temperature-doping phase diagram of Mo$_{1-x}$Ta$_x$Te$_2$ in Fig.~\ref{fig5}, which shows the Ta-doping dependence of $T_s$ and $T_c$. The structural transition temperatures acquired during warm-up ($T_{s, warm}$) and cool-down ($T_{s, cool}$) are defined by the extrema of the first derivative of $\rho(T)$ around the transition (see Fig.~S3 in Supplemental Material \cite{supp}). Both $T_{s,~warm}$ and $T_{s,~cool}$ show a generally decreasing trend with increasing $x$. Compared to $T_{s,~warm}$, $T_{s,~cool}$ decreases more rapidly with increasing $x$. When $x\geq 0.118$, both $T_{s,~warm}$ and $T_{s,~cool}$ are completely suppressed. On the other hand, after experiencing a local maximum at $x=$~0.042, $T_c$ also decreases when $x$ increases, and drops to zero at $x\geq 0.065$ (before $T_s$ vanishes). The disappearance of superconductivity is unique in our hole-doping phase diagram; in the previous phase diagram studies of MoTe$_2$ upon pressure \cite{Qi2016,Takahashi2017,Heikes2018,Lee2018,Guguchia2017,Hu2019}, isovalent chemical substitution \cite{Takahashi2017,Chen2016}, and electron doping \cite{Cho2017,Mandal2018}, they typically show the anticorrelation of $T_c$ and $T_s$ as well as a huge enhancement of $T_c$.

%%%%%%%%%%%%%%%%Figure 5
%%%%%%%%%%%%%%%%%%%%

To shed light on the issue of why the superconductivity of MoTe$_2$ is suppressed upon hole doping, a contour plot of $R_H$ is overlaid in Fig.~\ref{fig5}. We reveal that $\vert R_H \vert$ is significantly enhanced around the region when $T_s$ is suppressed to zero ($x\sim 0.1$), and $T_c$ vanishes when the enhancement of $\vert R_H \vert$ emerges at $x\sim 0.05$. Compared to the previous studies with other tuning parameters, while $T_c$ increases, low-temperature $\vert R_H \vert$ has either a weak electron-doping dependence \cite{Cho2017,Mandal2018} or decreases with pressure \cite{Hu2019}. Meanwhile, a similar enhancement of $R_H$ has been observed in another hole doping study, Nb-doped MoTe$_2$ \cite{Sakai2016}. Such enhancement is associated with the enhancement of thermopower divided by temperature $S/T$, which is maximum around the region where $T_s$ is completely suppressed; our $R_H$ contour plot is reminiscent of the contour plot of $S/T$ reported in the phase diagram of Nb-doped MoTe$_2$ (Fig.~1(b) in Ref.~\cite{Sakai2016}). According to Sakai \etal's argument, both enhancements of $R_H$ and $S/T$ are attributed to the strong fluctuation or phase separation around the nonpolar-polar structural transition, giving rise to some critical scattering effects on the carriers~\cite{Sakai2016}. Combining this statement with our phase diagram, the critical scattering may also hinder the formation of Cooper pairs, and therefore suppress superconductivity. Further investigations on the competition between superconductivity and critical scattering are highly desired to confirm this picture.

Another possible explanation for the suppression of superconductivity is related to the change in the Fermi surface topology upon hole doping. Cho \etal\ \cite{Cho2017} have performed theoretical calculations on the impact of electron and hole doping on $T_c$. While they have attributed the increase in $T_c$ upon electron doping (arising from Te vacancy in MoTe$_{2-x}$) to the enhancement of the density of states at the Fermi level ($N(E_F)$) and the electron-phonon coupling constant ($\lambda$), they have also predicted that, upon hole doping, $N(E_F)$ and $\lambda$ will be suppressed and therefore $T_c$ will decrease, which is consistent with our experimental findings. Cho \etal~further attributed the change in $\lambda$ to phonon vectors connecting between electron Fermi pockets, which are enlarged upon electron doping according to their calculations. In contrast, upon hole doping, electron pockets shrink and only spherical-shaped hole pockets remain at the $\Gamma$ point \cite{Sakai2016,Cho2017}. In the situation without phonon vectors linking between electron pockets, $\lambda$ will be suppressed and hence $T_c$ will be reduced. Therefore, our study has provided solid experimental evidence to showcase Cho \etal's theoretical prediction.

%%%%%%%%%%%%%%%%Figure 6
\begin{figure}[!t]\centering
       \resizebox{9cm}{!}{
              \includegraphics{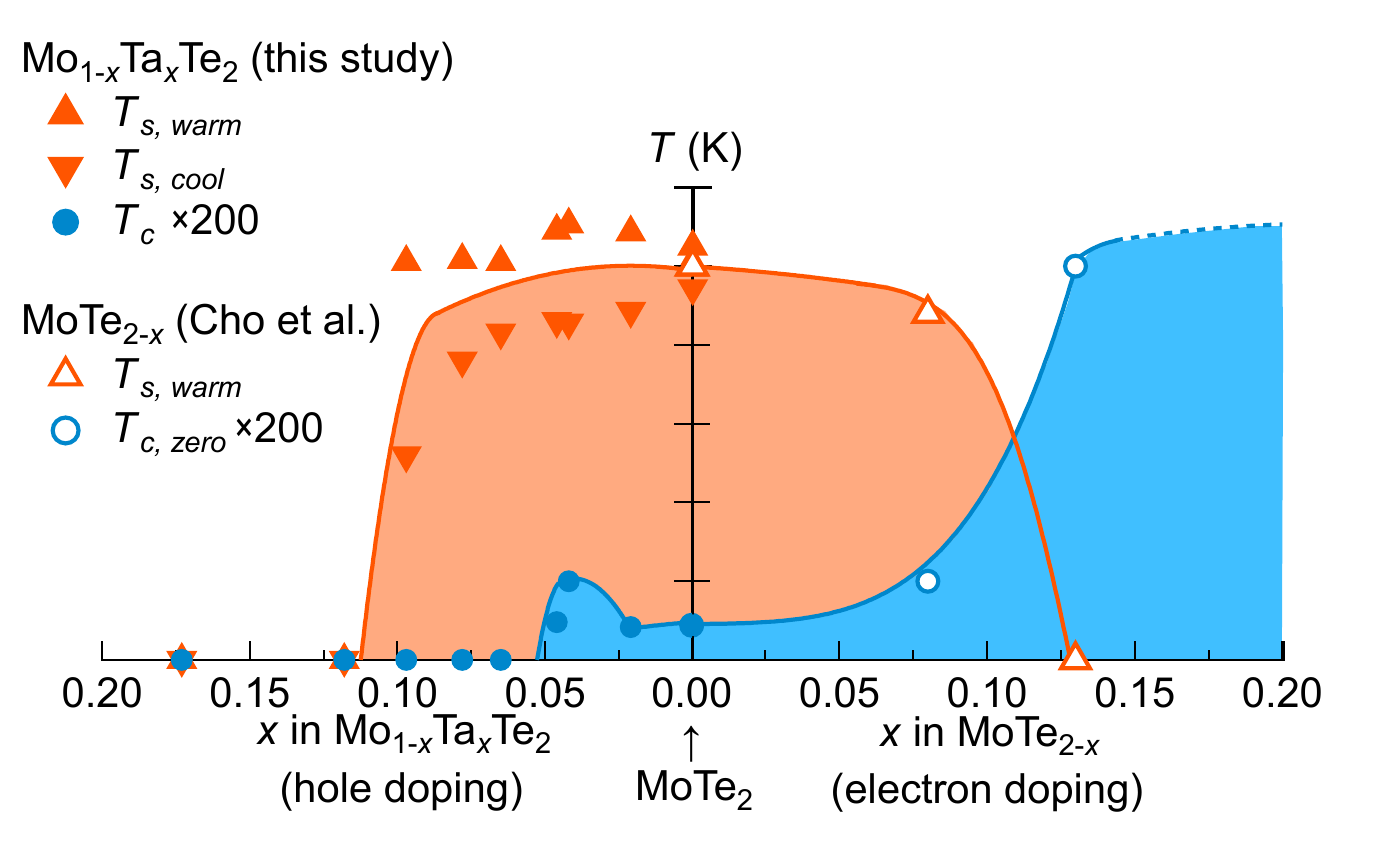}}                				
              \caption{\label{fig6} Temperature-doping phase diagram of hole-doped and electron-doped MoTe$_2$. The hole-doping data (solid symbols) are our findings of Mo$_{1-x}$Ta$_x$Te$_2$, as shown in Fig.~\ref{fig5}, while the electron-doping data (open symbols) are adapted from the findings of Te-deficient MoTe$_{2-x}$ from Cho \etal\ \cite{Cho2017}.
              }
              
\end{figure}
%%%%%%%%%%%%%%%%%%%%

To further elaborate on this idea, we connect our hole-doping phase diagram with the electron-doping phase diagram (based on the result of Te-deficient MoTe$_2$ from Cho \etal~\cite{Cho2017}) and plot the combined phase diagram in Fig.~\ref{fig6}. It unambiguously shows the asymmetry between the hole-doping phase and electron-doping diagrams, which is reminiscent of different behaviours between hole-doped and electron-doped cuprate superconductors~\cite{Armitage2010,Keimer2015}. While $T_s$ shows a similar suppression upon both hole- and electron-doping, the doping dependence of $T_c$ behaves differently. At the electron-doping region (the right-hand side of Fig.~\ref{fig6}), $T_c$ is largely enhanced. However, when we move to the hole-doping region (the left-hand side of Fig.~\ref{fig6}), $T_c$ is heavily suppressed. This demonstrates a clear trend that $T_c$ can be induced and enhanced when the electron carrier concentration increases, no matter what the phase is. 

Meanwhile, although the critical scattering around the structural transition may contribute to the suppression of superconductivity, our result shows that the tuning of the carrier concentration, which controls the phonon nesting vector(s), provides an effective means to vary the $T_c$ of MoTe$_2$, regardless of the suppression of $T_s$. These findings provide experimental evidence that enhancing the $T_c$ of MoTe$_2$ by solely increasing the electron carrier concentration while preserving the topologically nontrivial $T_d$ phase is possible. Such property can potentially boost the progress of the search for topological superconductivity in MoTe$_2$, which is currently hindered by its low $T_c$.

\section{Conclusions}

In summary, we have investigated the phase diagram of Ta-doped MoTe$_2$, Mo$_{1-x}$Ta$_x$Te$_2$, with $x=0-0.173$ through magnetotransport measurements. Single crystals of Mo$_{1-x}$Ta$_x$Te$_2$ were successfully grown by the self-flux method. X-ray diffraction and energy-dispersive X-ray spectroscopy have confirmed that Mo is partially substituted by Ta in the doped samples. By measuring the temperature dependence of resistivity and the Hall effect, we have revealed that the structural transition temperature $T_s$ is completely suppressed at $x\sim 0.11$, while the superconducting transition $T_c$ generally decreases upon Ta doping and finally vanishes at $x\sim 0.08$. This behaviour is in contrast to the previous phase diagrams constructed based on applying pressure, isovalent doping, or electron doping, which show the enhancement of $T_c$ when $T_s$ is suppressed. Moreover, the Hall coefficient is found to be enhanced at low temperatures around the region where $T_s$ is suppressed to zero, suggesting that the critical scattering arising from the structural temperature may have some contributions to the suppression of $T_c$. By comparing our findings with the phase diagram of electron-doped MoTe$_2$, we argue that the electron carrier concentration in MoTe$_2$ is a key factor in controlling $T_c$, which offers a straightforward way to boost the $T_c$ of MoTe$_2$.

Notes added: After the first submission of this article, we noticed a recently published article \cite{Zhang2023} which reports an enhancement of $T_c$ in Ta-doped MoTe$_2$. Our results do not agree with those of Ref. \cite{Zhang2023}. The discrepancy may be attributed to methodological differences. First, Ref. \cite{Zhang2023} used a different crystal growth condition. Second, we determine our $T_c$ values based on the observation of zero resistivity while Ref. \cite{Zhang2023} deduced their $T_c$ values from the onset of the transition in resistivity. We note that zero resistivity has not been observed in the doped samples in Ref. \cite{Zhang2023}.

\begin{acknowledgments}

We acknowledge Xinyou Liu, Ying Kit Tsui, Wei Zhang, and Lingfei Wang for fruitful discussions, and financial support from the Research Grants Council of Hong Kong (GRF/14300419, GRF/14301020 and A-CUHK402/19), CUHK Direct Grant (4053463, 4053528, 4053408 and 4053461), and the National Natural Science Foundation of China (12104384).

\end{acknowledgments}

%\bibliography{MoTe2}
%\bibliographystyle{apsrev4-1_local}

\providecommand{\noopsort}[1]{}\providecommand{\singleletter}[1]{#1}%

\end{document}